\documentclass[aps,10pt,twocolumn,letterpaper,english,preprintnumbers,amsmath,amssymb,superscriptaddress,footinbib,prx]{revtex4-2}

\usepackage{xcolor} 
\usepackage[colorlinks=true, linkcolor=blue, urlcolor=blue, citecolor=blue]{hyperref}

\usepackage{graphicx}
\usepackage[inkscapelatex=false]{svg}

\usepackage{amssymb,amsthm,amsmath}
\usepackage{xcolor,paralist,hyperref,titlesec,fancyhdr,etoolbox}
\usepackage{tikz}
\usepackage{tikz-qtree,tikz-qtree-compat}
\usepackage{cleveref}
\usepackage{float}     

\usepackage{tcolorbox} 
\usepackage{listings}  
\usepackage{xcolor}    

\tcbuselibrary{breakable,skins,listings}

\newtcolorbox[auto counter]{prompt}[2][]{
  label = #1,
  title={Prompt~\thetcbcounter\quad#2},
  breakable,
  enhanced,
  colback=white,
  listing only,
  listing options={
    basicstyle=\ttfamily,
    breaklines=true,
    language=Tex,
    escapechar=§,
  },
}

\lstdefinelanguage{json}{
  basicstyle=\normalfont\ttfamily,
  numbers=left,
  stepnumber=1,
  numbersep=5pt,
  showstringspaces=false,
  breaklines=true,
  frame=single,
  backgroundcolor=\color{white},
  literate=
   *{0}{{{\color{blue}0}}}{1}
    {1}{{{\color{blue}1}}}{1}
    {2}{{{\color{blue}2}}}{1}
    {3}{{{\color{blue}3}}}{1}
    {4}{{{\color{blue}4}}}{1}
    {5}{{{\color{blue}5}}}{1}
    {6}{{{\color{blue}6}}}{1}
    {7}{{{\color{blue}7}}}{1}
    {8}{{{\color{blue}8}}}{1}
    {9}{{{\color{blue}9}}}{1}
    {:}{{{\color{red}{:}}}}{1}
    {,}{{{\color{red}{,}}}}{1}
    {\{}{{{\color{black}{\{}}}}{1}
    {\}}{{{\color{black}{\}}}}}{1}
    {[}{{{\color{black}{[}}}}{1}
    {]}{{{\color{black}{]}}}}{1},
}

\newtcolorbox{userinput}{
  breakable,
  enhanced,
  colback=blue!5,
  colframe=blue!75!black,
  title=User Input,
}

\titleformat{\section}[display]{\normalfont\huge\bfseries\centering}{\centering\chaptertitlename\thechapter}{10pt}{\Large}
\titlespacing*{\section}{0pt}{0ex}{0ex}

\hypersetup{ colorlinks=true, linkcolor=black, filecolor=black, urlcolor=black }

\DeclareMathAlphabet\mathbfcal{OMS}{cmsy}{b}{n}

\newcommand{\ourtitle}{Random Tree Model of Meaningful Memory}
 \newcommand{\iasaff}{School of Natural Sciences, Institute for Advanced Study, Princeton, NJ, 08540, USA}
  \newcommand{\emory}{Department of Physics, Emory University, Atlanta, GA 30322, USA}
 \newcommand{\weizmann}{Department of Brain Sciences, Weizmann Institute of Science, Rehovot, 76100, Israel}

\begin{document}
\title{Random Tree Model of Meaningful Memory}
\author{Weishun Zhong}
\affiliation{School of Natural Sciences, Institute for Advanced Study, Princeton, NJ, 08540, USA}

\author{Tankut Can}
\affiliation{School of Natural Sciences, Institute for Advanced Study, Princeton, NJ, 08540, USA}
\affiliation{Department of Physics, Emory University, Atlanta, GA 30322, USA}

\author{Antonis Georgiou}
 \affiliation{School of Natural Sciences, Institute for Advanced Study, Princeton, NJ, 08540, USA}
 \affiliation{Department of Brain Sciences,
Weizmann Institute of Science, Rehovot, 76100, Israel}

\author{Ilya Shnayderman}
 \affiliation{Department of Brain Sciences,
Weizmann Institute of Science, Rehovot, 76100, Israel}

\author{Mikhail Katkov}
 \affiliation{School of Natural Sciences, Institute for Advanced Study, Princeton, NJ, 08540, USA}
 \affiliation{Department of Brain Sciences,
Weizmann Institute of Science, Rehovot, 76100, Israel}

\author{Misha Tsodyks}
\email{Corresponding author: mtsodyks@gmail.com}
 \affiliation{School of Natural Sciences, Institute for Advanced Study, Princeton, NJ, 08540, USA}
 \affiliation{Department of Brain Sciences,
Weizmann Institute of Science, Rehovot, 76100, Israel}

\date{\today}

\begin{abstract}
Traditional studies of memory for meaningful narratives focus on specific stories and their semantic structures but do not address common quantitative features of recall across different narratives. We introduce a statistical ensemble of random trees to represent narratives as hierarchies of key points, where each node is a compressed representation of its descendant leaves, which are the original narrative segments. Recall is modeled as constrained by working memory capacity from this hierarchical structure. Our analytical solution aligns with observations from large-scale narrative recall experiments. Specifically, our model explains that (1) average recall length increases sublinearly with narrative length, and (2) individuals summarize increasingly longer narrative segments in each recall sentence. Additionally, the theory predicts that for sufficiently long narratives, a universal, scale-invariant limit emerges, where the fraction of a narrative summarized by a single recall sentence follows a distribution independent of narrative length.

\end{abstract} 

\maketitle

\let\thefootnote\relax

\bigskip

Narratives are a naturalistic form of stimuli for probing the structure and organization of human cognitive functions \cite{thorndyke1977cognitive, bower1976experiments,lee2020can,hasson2015hierarchical,lerner2011topographic,bietti2019storytelling}. In the classic work of Bartlett \cite{bartlett1995remembering}, a close link between narrative memory and comprehension was established, and both were shown to vary greatly across individuals. Subsequent studies established that narrative comprehension and recall is strongly linked to a subject's semantic knowledge, such as schemas, that help to interpret new information in the context of prior knowledge (see e.g. \cite{alba1983memory, baldassano2018representation, rumelhart1975notes, schank2013scripts}). It appears therefore that the complexity of memory for meaningful information cannot be captured by simple physics-style models with a few general postulates and mathematical tractability. Indeed, in the psychological literature, most studies aim to relate memory for narratives to their linguistic organization and do not predict generic quantitative features of narrative recall (see e.g. \cite{thorndyke1980critique,kintsch1998comprehension}), in contrast to studies of memory for random material where many quantitative experimental and theoretical results are obtained (see e.g. \cite{kahana2020computational}). In particular, some of the authors previously developed a model that predicted the universal relation between the average number of recalled items and the number of items retained in memory after list presentation, and this relation was well supported by experimental results \cite{naim2020fundamental}. As opposed to random lists, narratives have a complex structure at different levels of abstraction, and their recall cannot be evaluated simply by counting the number of recalled words because people do not recall meaningful material verbatim; moreover, the recall length alone does not necessarily reflect its quality because concise summary of the narrative could better communicate its meaning than longer but less structured recall. A theory should capture these complexities in order to be relevant for understanding the memory of meaningful material. The model presented below is motivated by our recent
large-scale experiments 
with narratives of various lengths \cite{georgiou2023using}. Unsurprisingly, we observed that recall is qualitatively different from that of random lists: 
people tend to recall information from the narrative in the same order as it is presented \cite{trabasso1985causal}, as opposed to much more variable recall order of random lists \cite{kahana1996associative,tulving1962subjective, howard1999contextual}; for longer narratives, people compress progressively larger pieces of information from the narrative into single sentences, resulting in a sublinear growth of recall length with the narrative size \cite{georgiou2023using}.

\begin{figure*}[ht]
    \centering
    \includegraphics[width=1\textwidth]{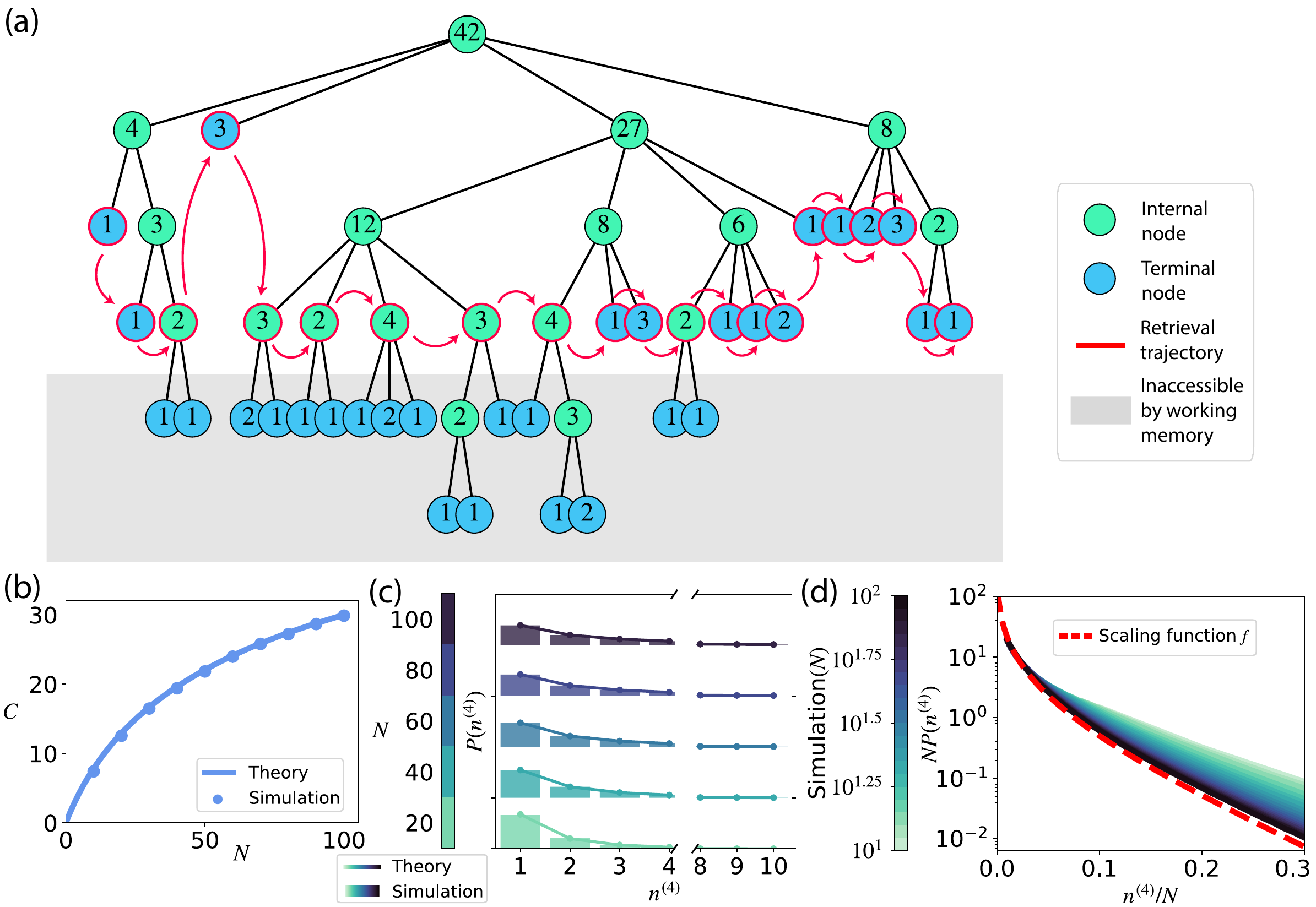}
    \caption{\textbf{Ensemble of random trees.} \\
    \textbf{(a)} Schematics of memory retrieval from a random hierarchical representation. An example of a single realization of the random tree created by the model for $N=42$ encoded clauses with branching ratio $K=4$ (empty nodes are not shown). Internal nodes are shown in green, while the terminal nodes (leaves) are shown in blue. The tree generating process starts with the whole narrative contained in the root node (level 1), which is subsequently split into up to 4 chunks at the next level. The splitting continues self-similarly until either a chunk fails to split further (e.g., the blue ``4'' at level 3) or becomes a single clause (the blue ``1''s). The grey shaded area illustrate the limit imposed by working memory capacity as retrieval starts by descending from the root (retrieved nodes are shown in red). (b)-(c) Comparison between analytical solution and numerical simulations. \textbf{(b)} Mean recalled length $C$ as a function of encoded length $N$. Numerical simulations are averaged over $10^4$ realizations of random trees (see details in SI Sec.~A). \textbf{(c)} Distribution of chunk size at the $D^{th}$ level $n^{(D)}$ ($D=4$), given root size $n^{(1)}=N$, range between tick marks in the y-axis corresponds to $[0,1]$. \textbf{(d)} Distribution of compression ratios scaled by $N$ as a function of the compression ratios divided by $N$. Simulations of different $N$ are shown in different shades of green. The red dashed line is the asymptotic scaling function from Eq.~\eqref{eq:scaling_function}.  }
    \vspace{-1em}
\label{fig:schematics}
\end{figure*}
In the current study, we show that despite the more complex nature of meaningful recall, the above statistical regularities can be captured by mathematical modeling. To this end, we propose a simplified model of narrative recall that focuses on describing the statistical aspects of recall across large populations of people rather than addressing individual differences between people and narratives.

\begin{figure*}[ht]
    \centering
    \includegraphics[width=0.9\linewidth]{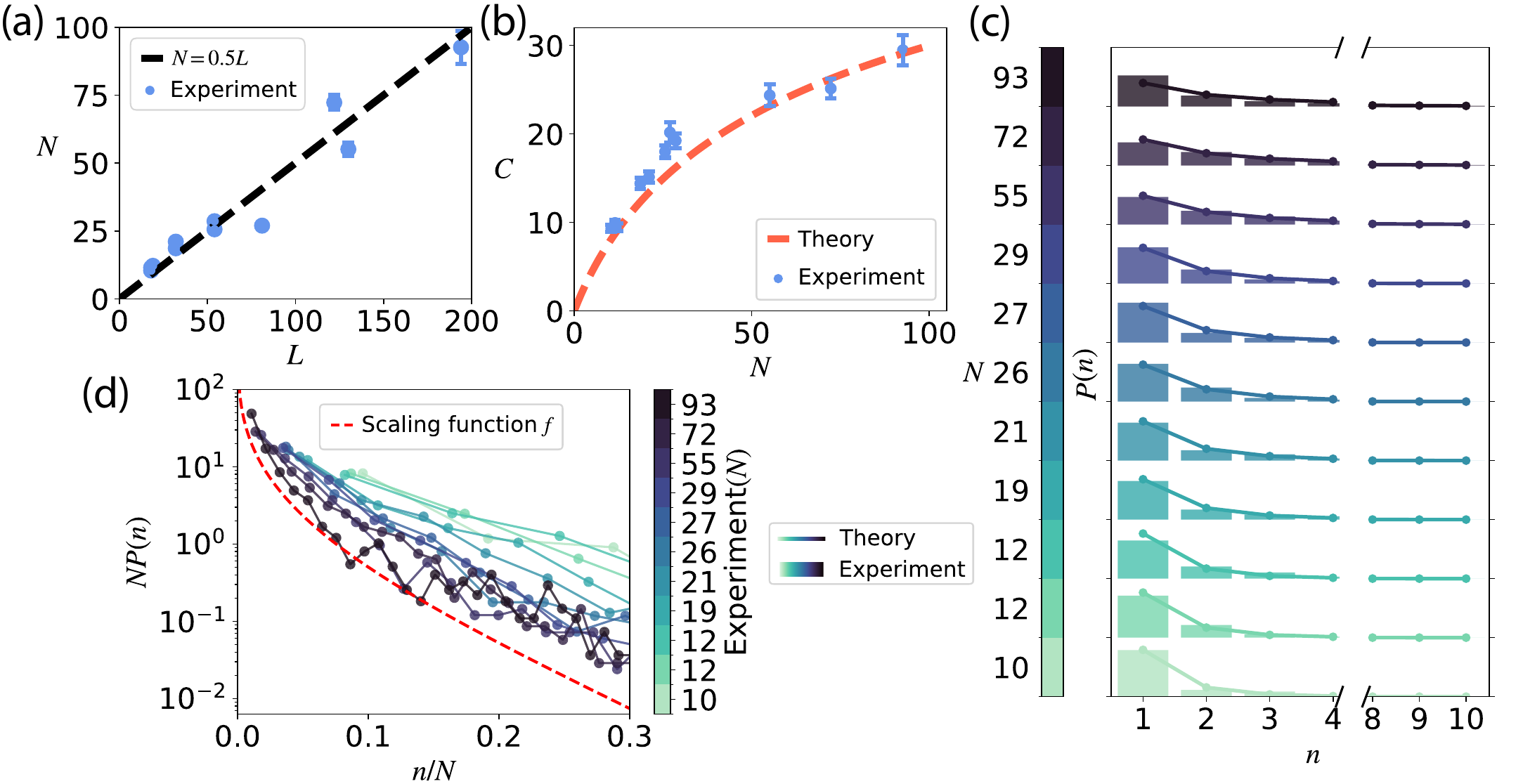}
    \caption{{\bf Comparison between theory and experiment.} \textbf{(a)} Average size of the tree memory representation of each narrative ($N$), estimated as explained in the text, plotted vs narrative length ($L$) for 11 narratives in the dataset. Dashed line corresponds to $N=0.5L$. \textbf{(b)} The mean number of recalled clauses $C$ vs average $N$, for all 11 narratives. Blue filled circles - data. Red dashed line - theoretical prediction obtained from Eqs.~\eqref{eq:Pn}-\eqref{eq:CvsK,D} with $K=D=4$. Error bars in (a,b) are standard error of the mean. \textbf{(c)} Normalized empirical histograms of compression ratios for all subjects separately for each narrative, as measured from mapping recalled clauses back to the narrative clauses. Data for different narratives are shown in color corresponding to the colorbar marked with values of $N$ for each narrative. Solid lines - theoretical predictions obtained from Eq.~\eqref{eq:Pn} with $K=D=4$. Range between tick marks in y-axis is $[0,1]$. {\bf (d)} The distribution of experimentally measured compression ratios relative to $N$ approaches the universal scale-invariant scaling function $f$ in Eq.~\eqref{eq:scaling_function} as $N$ increases. }
    \vspace{-1em}
    \label{fig:data_figure}
\end{figure*}
Our model is based on two basic principles. The first principle describes the memory encoding of a narrative. Following multiple previous studies, we consider a narrative as a linear sequence of clauses, i.e., the shortest meaningful pieces of text (see e.g. \cite{labov2013language}). We assume, however, that after people {\it comprehend} the narrative, they form a tree-like memory representation of it where, at each level of the tree, information units of a given level of abstraction are encoded \cite{kintsch1978toward, mandler1977remembrance, gernsbacher2013language}, with each unit representing a summary of a specific segment of the narrative. In particular, the root of the tree represents the whole narrative; the upper-level nodes represent pieces of the narrative that correspond to its main keypoints \cite{bar-haim2020-arguments}; and lower levels represent shorter pieces corresponding to progressively finer points \cite{graesser1997discourse, van1983strategies, lorch1985topic, cattan2023-key}. Moreover, we assume that a piece corresponding to a particular keypoint is split into pieces corresponding to the finer points that elaborate the original keypoint, resulting in a tree structure of a narrative. We see the evidence for this type of memory representation in the fact that people can easily summarize a familiar narrative at different levels of detail, which we believe is achieved by retrieving the node representations at different tree levels, e.g., retrieving upper nodes evokes the most abstract memory of a narrative in terms of its most general points. An implicit assumption made here is that continuous stimuli, such as a text, is represented by a discrete set of nodes. This is also supported by experiments that show humans naturally segment sequential stimuli (e.g. movies, personal experiences) into discrete units \cite{radvansky2017event, zacks2001human,ben2018hippocampal,kalm2012neural,baldassano2017discovering,zheng2022neurons}. There is also extensive support for the claim that narratives, from folk tales \cite{colby1973partial} to oral narratives \cite{Labov1966,labov2013language}, have such a hierarchical, tree-like structure. We argue that this hierarchy is reflected in the mental representation. Finally, as we observed in our previous publication \cite{georgiou2023using}, people only remember a fraction of the narrative clauses as attested by recognition experiments; besides, it seems plausible that not all recognizable clauses may be integrated into the tree representation. We therefore leave the size of the tree, $N$, as a subject-dependent parameter that should be estimated from the data, as we show below. 

The second principle reflects the way people recall a given narrative based on its memory representation described above. We assume that people mentally traverse the tree from upper to lower levels, keeping the intermediate nodes in working memory, to constantly maintain the integrity of the narrative in mind. It is in this respect that we believe the recall of meaningful narratives is fundamentally different from the recall of random lists of words. In particular, working memory capacity limits the number of nodes that can simultaneously be kept in mind, thus limiting the number of levels that can be reached during recall. We further assume that when reporting a particular node during recall, people use a single clause to express the content of the corresponding piece.

It seems plausible that a specific instantiation of the memory tree depends both on the narrative structure and on the way a particular person comprehends and subsequently memorizes it. Hence, in order to translate the above two principles into a quantitatively constrained model, we make a highly simplifying assumption about the statistical ensemble of trees encoding different narratives of a particular length in the minds of different people. We build this ensemble by a recursive self-similar process in which a narrative is split into up to $K$ pieces by the random placement of $K-1$ boundaries, and each piece is similarly split into up to $K$ pieces, and so on. We begin this process from the whole narrative and continue it recursively from one layer to the next until splitting stops (see details in SI Sec.~A). The value of $K$ could be considered a free parameter to be determined from the results. We assumed however that $K$ is related to working memory capacity. Indeed, people can effortlessly summarize a narrative at the highest level of abstraction, which in our scheme would correspond to activating the representations of the upper-level keypoints. Therefore, we assumed that narrative representations are constructed in such a way that summarizing the whole narrative (or any of its meaningful segments) does not overwhelm working memory capacity and therefore chose $K=4$  \cite{cowan2010magical}.  

We then illustrate the recall process playing out on this tree, assuming that people systematically traverse the memory representation by following each route beginning from the root, until they either reach a leaf or a lowest reachable ($D^{th}$) level (the levels above the grey shaded area in Fig.~\ref{fig:schematics}(a)). We again consider $D=4$ because it corresponds to working memory capacity. The retrieval trajectory is shown in red on Fig.~\ref{fig:schematics}(a). One can see that some of the nodes reached by a retrieval trajectory contain single narrative clauses, while others contain multiple clauses (up to 12 for the illustrated tree). Since we assume that each node, no matter its level, is recalled with a single recall clause, the length of recall is defined by the number of retrieved nodes, while each recall clause summarizes the corresponding number of narrative clauses. It is important to emphasize that while the ensemble of memory representations described above involves a random set of trees, the hypothesized recall process itself is deterministic, i.e. the number of recall clauses and the pieces of the narrative that each one of them corresponds to are fully determined by the memory representation of the narrative. 

The model described above can be simulated numerically and allows for an asymptotic mathematical solution as we show below. The length of recall (the number of clauses in recall, $C$) averaged over the ensemble of trees with $N$ clauses exhibits a characteristic saturating form for increasing $N$, see Fig.~\ref{fig:schematics}(b). We also computed the distribution of the number of narrative clauses summarized by single recall clauses (i.e., compression ratios of different recall clauses), shown in Fig.~\ref{fig:schematics}(c) for compression ratios up to 10 and for different $N$. One can see that, as expected, for longer narratives compression ratios of recall clauses tend to increase. 

To obtain the analytical approximations for these numerical results, we derive a recurrence relation for the distribution of the number of clauses represented by nodes at a given level $l$, $P(n^{(l)})$, with
\begin{equation}
   P(n^{(1)}) = \delta(n^{(1)} - N),
\end{equation}
since the upper node represents all of the narrative clauses retained in memory after acquisition. When a node at level $l$ with $n^{(l)}$ clauses is split randomly into $K$ nodes of level $l+1$ (some of them empty), the distribution of the size of each of the resulting nodes can be computed by the ``stars and bars'' method (\cite{flajolet2009analytic}) by considering all possible configurations of $n^{(l)}$ stars and $K-1$ bars and only counting the number of stars to the left of the left-most bar. This results in the following recursive expression (for $n^{(l+1)}\leq n^{(l)}$, 0 otherwise):
\begin{align}
    P(n^{(l+1)}|n^{(l)}) &= \frac{Z_{K-1}(n^{(l)}-n^{(l+1)})}{Z_K(n^{(l)})},\label{eq:markov_trans} \\
    Z_K(n):&= \binom{n+K-1}{K-1}.
\end{align}
Applying this recursive expression and integrating over intermediate levels results in the distribution of the node sizes in level $D$,
\begin{equation}
    P(n^{(D)}) = \sum_{{\substack{0\leq n^{(D-1)} \\ \leq \dots \leq n^{(1)}}}}
    \prod_{l=1}^{D-1} P(n^{(l+1)}|n^{(l)})P(n^{(1)}).
    \label{eq:Pn}
\end{equation}
Note that Eq.~\eqref{eq:Pn} defines a Markov chain, see SI Sec.~B for more details. In this version of the stars and bars model, some of the nodes may end up being empty (if two or more bars are adjacent), and the average number of nonempty nodes at level $D$ will define the average length of recall, $C= K^{D-1}[1-P(n^{(D)}=0)]$, which is found to be (see SI Sec.~B for derivations and its connection with the Riemann zeta function):
\begin{align}
C = K^{D-1} \sum_{m = 1}^{N} \left( { N \atop m}\right)\frac{(-1)^{m+1}}{ Z_{K}(m)^{D-1}} , \label{eq:CvsK,D}
\end{align}
while the re-normalized distribution of non-zero node sizes corresponds to the distribution of compression ratios introduced above. As $N$ increases, the probability to have an empty node approaches zero, and this equation therefore predicts that $C$ saturates at $K^{D-1}$ (i.e. $64$ for our choice of $K=D=4$) for long narratives. For $K=D=4$, the  analytical results obtained with Eqs.~\eqref{eq:Pn}-\eqref{eq:CvsK,D} are very close to numerical simulations, as shown in Fig.~\ref{fig:schematics}(b,c). In particular, one can see from Fig.~\ref{fig:schematics}(c) that as the narrative gets longer, the distribution of compression ratios shifts to the right, i.e. recall clauses tend to summarize progressively bigger pieces of the narrative. 

For long narratives, the distribution of compression ratios, normalized by narrative size ($s = \frac{n}{N}$), asymptotes to a universal scaling function $f(s)$ (see derivations in SI Sec.~B):
\vspace{-0.5em}
\begin{align}
    &P(n^{(D)}) = \frac{1}{N} f(n^{(D)}/N), \label{eq:asymp_Pn}\\
       &f(s_D) = \prod_{l=1}^{D-1} \int_{s_{l+1}}^1 ds_l \rho(s_{l+1}|s_l) \rho(s_{1}|1), \label{eq:scaling_function}\\
    &\rho(s_{l+1}|s_l):= \frac{K-1}{s_l} \left(1-\frac{s_{l+1}}{s_l} \right)^{K-2}.
\end{align}

In this limit, $f(s)$ is a scale-invariant probability density function for the normalized compression ratio $s$, which in particular implies that the average compression ratios grow proportionally to narrative length. The shape of the scaling function computed with Eq.~\eqref{eq:scaling_function} with $K=D=4$ is shown in Fig.~\ref{fig:schematics}(d) in red, together with finite-$N$ simulations from the random tree model. 

We now compare our theoretical predictions to experimental data on 11 narratives ranging from 19 to 194 clauses in length, including 8 narratives from our previous study \cite{georgiou2023using} and 3 new narratives selected for this study. All of the narratives, except for two, were chosen from the study by Labov \cite{labov2013language} or generated by a Large Language Model (LLM), GPT-4 from OpenAI \cite{achiam2023gpt}, using Labov's narratives as templates. The original narratives were edited for clarity and spelling. The two remaining narratives were borrowed from \cite{johnstone1990stories}. We used the clause segmentation from \cite{labov2013language} for narratives sourced from that work, and this clause segmentation was also transferred to GPT4-generated narratives that used the originals as a templates. Each narrative was presented to 100 subjects on their computer screens via the internet platform Prolific (www.prolific.com), who were instructed to recall it as closely as possible to the original. We followed the free recall experimental protocol described in detail in \cite{georgiou2023using}. We used GPT-4 to segment the recalls into single clauses, using the same prompts as in \cite{georgiou2023using} (reproduced in SI Sec.~C). To obtain compression ratios for all recall clauses, we instructed GPT-4 to map each recall clause back to the original narrative and then counted the number of narrative clauses it was mapped to (see SI Sec.~D for details of this analysis). We reason that the narrative clauses to which a given recall clause is mapped correspond precisely to the segment of the narrative encoded by the corresponding node, which was recalled by this clause. The crucial remaining step needed to compare these results to the theoretical predictions above is to estimate the number of clauses in the memory representation of a given narrative for each subject ($N$). We compute this estimate as the total number of narrative clauses into which at least one recall clause is mapped. We assume that the remaining narrative clauses are either not remembered by the subject, or are not integrated into the tree memory representation of the narrative and hence cannot be recalled. The average values of $N$ obtained for each narrative are plotted against their lengths in Fig.~\ref{fig:data_figure}(a) and are observed to be close to half the total length of the narrative. With this assumption, experimental results for the average length of recall and compression ratios up to 10 are quite similar to theoretical predictions, as shown in Fig.~\ref{fig:data_figure}(b,c), even though no parameters were tuned after the fact to fit the model predictions to the data. Fig.~\ref{fig:data_figure}(d) shows that for longer narratives, the distribution of compression ratios relative to $N$ indeed approaches the scale-invariant form predicted by Eqs.~\eqref{eq:asymp_Pn}-\eqref{eq:scaling_function}, with the speed similar to corresponding simulations in the same range of $N$s shown in Fig.~\ref{fig:schematics}(d).

It is important to point out that using the compression ratios as a quantitative characterization of recall, while intuitively appealing, is not devoid of challenges. Since it relies on mapping individual recalls to original narratives, it is quite subjective and extremely time-consuming, necessitating the usage of language models like GPT-4 \cite{achiam2023gpt}. Random checks by the authors showed that humans agree well on clauses with small compression ratios (more concrete ones) but disagreement can be substantial for clauses with large compression ratios (more abstract ones). This could be because mapping abstract clauses requires a high degree of understanding of both the narrative and the recall, which according to our model implies constructing corresponding tree representations and matching them to each other, both exhibiting individual differences between people. To check how robust the mappings are when performed by language models, and how robust the statistical features of recalls predicted by the model are, we repeated the mapping with the recently released DeepSeek model \cite{liu2024deepseek}. Comparing the mappings on a clause-by-clause basis indeed revealed that they are not always aligned, with the differences becoming more apparent for longer narratives, especially for recall clauses with high compression ratios. Despite this variability, statistical features of recall, namely the distributions of compression ratios over recall clauses, and the dependence of the average recall length on the estimated average size of the tree representations, are similar to our theory's predictions when using the mappings generated by both models (SI Sec.~F).  


{\bf In summary}, we showed that despite its highly complex nature, some statistical trends in meaningful narrative recall can be captured by a mathematical model based on two basic principles: random tree-like memory representations and a working memory-limited deterministic recall process. In particular, the model correctly predicts the average length of recall and the distribution of compression ratios over recall clauses as a function of the size of the narrative tree representation. Since we do not have a theoretical way to calculate the latter for each narrative/subject, we estimated the size of the tree using LLM-performed mapping of each recall back into the narrative. We found that the average tree size of a narrative, estimated this way, is close to one-half of the total number of clauses in a narrative for our set of narratives. Note that this fraction is lower than that of the clauses retained in memory, which was estimated from recognition experiments in \cite{georgiou2023using} to be about 70\%, indicating that some fraction of retained clauses is not integrated into the tree representation. We currently do not have a theoretical understanding of this observation and it would be interesting to study whether this simple relation holds for narratives of different types. Another important issue not addressed in our model is how memory representations are formed during and immediately after acquisition. We believe that the encoding tree is built gradually as subjects comprehend the narrative, which will have to be studied in future experiments and models. The same could be said about narrative \textit{production}, which in our view involves recalling autobiographical events while also adding other information needed to make the recall understandable.

All of the narratives chosen for this study are of the same type. Therefore, it would be interesting for future work to study whether and how our model generalizes to other types of narratives, such as fictional ones.

The model has two parameters: the maximal splitting ratio in the random tree ensemble of narrative representations and the maximal depth of the recall process. Importantly, these parameters were not fine-tuned to match the model and the data but rather both were taken to be equal to human working memory capacity. The model hence suggests the critical role of working memory capacity in controlling narrative recall. This novel suggestion should be further investigated in future studies. 

The narrative encoding scheme hypothesized in this study is clearly very simplified, and real schemes must contain multiple deviations from a strict hierarchy, such as, e.g., direct causal links between different keypoints. The effects of these non-hierarchical encoding features could be studied at the level of individual recalls; however, we believe that, as the results presented in the study show, they are relatively minor at the statistical level.

Finally, the self-similarity of tree representations suggested in this study resembles the statistical similarity in autobiographical recall at different time scales observed in \cite{maylor2001scale}. It will be interesting to pursue this analogy and investigate further whether the statistics of recall of long narratives at different levels is also similar. In particular this would imply that as narratives get longer, the average recall length saturates, with each clause becoming more and more abstract, summarizing progressively longer parts of the narrative.
 
\acknowledgments{
The authors acknowledge Doron Sivan for assistance in validating the recall mappings. W.Z. is supported by Eric and Wendy Schmidt Member in Biology Fund and the Simons Foundation at the Institute for Advanced Study. A.G. is supported by the Martin A. and Helen Chooljian Member in Biology Fund and the Charles L. Brown Member in Biology Fund. MK is supported in part by a grant from Fran Morris Rosman and Richard Rosman. M.T. is supported by the Simons Foundation, MBZUAI-WIS Joint Program for Artificial Intelligence Research and Foundation Adelis. }

\bibliographystyle{unsrt}
\bibliography{ref}

\clearpage

\pagebreak

\setcounter{page}{1}
\setcounter{equation}{0}
\setcounter{figure}{0}
\renewcommand{\theequation}{S.\arabic{equation}}
\renewcommand{\thefigure}{S\arabic{figure}}
\renewcommand*{\thepage}{S\arabic{page}}

\renewcommand{\sectionmark}[1]{}
\renewcommand{\subsectionmark}[1]{}

\onecolumngrid

\begin{center}
{\large \textbf{Supplementary Information for \\``\ourtitle"}}\\
\vspace{0.25cm}

Weishun Zhong$^{1}$,  Tankut Can,$^{1,2}$ Antonis Georgiou,$^{1,3}$ Ilya Shnayderman$^{3}$, Mikhail Katkov,$^{1,3}$ and Misha Tsodyks$^{1,3}$

\textit{
$^1$\iasaff\\
$^2$\emory \\
$^3$\weizmann\\
}

\end{center}

\subsection{Random tree ensemble and numerical simulations}

The ensemble of random trees is built by a self-similar splitting process: it begins with the root node that contains all $N$ clauses and progressively splits it into smaller and smaller nodes at lower levels. At each level $l$, a node of size $n^{(l)}$ randomly splits into $K$ nodes at level $l+1$, of sizes $n^{(l+1)}_{i} \geq 0$, $i=1,\dots,K$, such that
\begin{equation}
\label{eq:split}
    \text{(split)}\qquad n^{(l)} = \sum_{i=1}^K n^{(l+1)}_{i}.
\end{equation}
This is achieved by randomly inserting $K-1$ dividing bars among $n^{(l)}$ clauses. More precisely, $n^{(l)}$ clauses and $K-1$ divisions are randomly placed in $n^{(l)} + K - 1$ positions, and $n^{(l+1)}_i$ is defined as the number of clauses to the left of the first bar for $i=1$, between $(i-1)^{th}$ and $i^{th}$ bars for $i=2,...,K-1$ and to the right of the last bar for $i=K$. 
A child node with $n^{(l+1)}_{i}$ clauses stops splitting further, and becomes a leaf, if
\begin{equation}
\label{eq:stop}
    \begin{split}
        \text{(stop)}\qquad n^{(l+1)}_i &= n^{(l)}\;\; \text{or} \\
    n^{(l+1)}_i &\leq 1.
    \end{split}
\end{equation}
The first stopping condition is triggered when a node at the $l^{th}$ level fails to split further at the $(l+1)^{th}$ level, and is motivated by the idea that when comprehending a narrative, individuals may encode certain events as coherent chunks rather than further segmenting them into finer details—a phenomenon commonly observed in memory experiments where subjects employ chunking strategies, e.g., \cite{simon1974big}. The second stopping condition follows from the assumption that single clauses are considered the smallest meaningful units of a narrative. The entire splitting process ends once all the children nodes have met the stopping condition. 

To simulate recall, as described in the main text, we cut off all the nodes below the $D^{th}$ level and retrieve all the leaves of the remaining tree. The size of a retrieved node is then considered to be the compression ratio for the corresponding recall clause, as explained in the paper. To obtain the numerical results presented in Fig.~\ref{fig:schematics}(b)-(c), for each value of $N$, 10,000 random tree realizations were constructed with the splitting parameter $K=4$, and for each tree, recall was simulated with $D=4$. For the results shown in Fig.~\ref{fig:schematics}(d), 20 different values of $N$ were sampled uniformly on a log scale from $N=10$ to $N=100$, and $10,000N$ random realizations were simulated for each $N$ in order to obtain good statistics for the tail of the compression ratios distribution.

\subsection{Analytical solution of the model}

Mathematically, the splitting process described above is akin to the problem of the weak composition of an integer $n$ into $K$-tuples of nonnegative integers, a process called ``Stars and Bars'' \cite{flajolet2009analytic} (S\&B), which counts all possible configurations of splitting $n$ ``stars'' into $K$ bins by placing $K-1$ ``bars'' among them. However, the S\&B process is only an approximation of the tree generation process described in Eq.~\eqref{eq:split}-\eqref{eq:stop} because of the first stopping condition in Eq.~\eqref{eq:stop}. In S\&B, a node stops splitting only when it becomes non-splittable, i.e., when the second condition in Eq.~\eqref{eq:stop} is met. However, as $n^{(l)}$ becomes large, the probability of having $n^{(l+1)}_i = n^{(l)}$ becomes increasingly small, and the approximation using S\&B converges to the exact distribution. In fact, for the range of $N$ considered in this paper (main text Fig.~\ref{fig:schematics}-\ref{fig:data_figure}), there is little difference between the theory and the simulation, and the approximation is nearly exact. It is in this sense that we refer to the solution obtained below using S\&B as an analytical solution to Eq.~\eqref{eq:split}-\eqref{eq:stop}.

The total number of weak compositions is given by the S\&B Theorem 2 \cite{flajolet2009analytic}, 
\begin{equation}
\label{eq:binom}
    Z_K(n) = \binom{n+K-1}{K-1},
\end{equation}
which states that there are $Z_K(n)$ ways to partition $n$ indistinguishable objects (stars) into $K$ distinguishable adjacent bins (whose boundaries correspond to bars), by placing $K-1$ bars randomly among the $n+K-1$ total positions, where each position can either be occupied by a star or a bar.

Now consider splitting $n^{(l)}$ stars into $K$ bins. Without loss of generality, let's focus on the first bin. The number of configurations with the first bin having $n^{(l+1)}$ stars is $Z_{K-1}(n^{(l)} - n^{(l+1)})$, since there are $n^{(l)} - n^{(l+1)}$ remaining stars that have to be divided into $K-1$ remaining bins. Therefore, the probability of observing $n^{(l+1)}$ stars in the first bin of the $(l+1)^{th}$ level is given by
\begin{equation}
\label{eq:conditional}
P(n^{(l+1)}|n^{(l)}) = \frac{Z_{K-1}(n^{(l)}-n^{(l+1)})}{Z_K(n^{(l)})}, 
\end{equation}
The same equation holds for other bins as well, not just the first one. For example, the number of configurations with $n^{(l+1)}$ stars in the second bin can be obtained by gluing these stars with the two surrounding bars and considering this unit as a new bar, thereby reducing the problem again to the situation with $n^{(l)} - n^{(l+1)}$ stars and $K-1$ bars. We are interested in computing the distribution of node sizes (or compression ratios) at level $D$, $p(n^{(D)})$, which can be obtained by repeatedly applying Eq.~\eqref{eq:conditional} starting from level 1: 
\begin{align}
    P(n^{(D)}) &= \sum_{{\substack{0\leq n^{(D-1)} \\ \leq \dots \leq n^{(1)}}}}
    \prod_{l=1}^{D-1} P(n^{(l+1)}|n^{(l)})P(n^{(1)})  \label{appeq:marginal}\\
    &= \sum_{n^{(D-1)}} P(n^{(D)}|n^{(D-1)}) P(n^{(D-1)}), \label{appeq:markov_chain} \\
    P(n^{(1)}) &= \delta(n^{(1)}-N).
\end{align}
Eq.~\eqref{appeq:markov_chain} defines a Markov chain $\{n^{(1)}, n^{(2)},\dots, n^{(D)}\}$ with transition probabilities given by $P(n^{(l+1)}|n^{(l)})$. We can define the state vector as the probability distribution of compression ratios at the $l^{th}$ level, 
\begin{align}
\label{eq:state_vector}
    \mathbf{P}^{(l)}:= \left(P(n^{(l)}=0),P(n^{(l)}=1), \dots, P(n^{(l)}=N)\right)^{\mathbf{T}}
\end{align}
and the transition matrix as the conditional probabilities in Eq.~\eqref{eq:conditional},
\begin{align}
\label{eq:transition_matrix}
    \left(\mathbfcal{T}\right)_{ij}:= P(n^{(l+1)}=i|n^{(l)}=j).
\end{align}
In particular, the transition matrix satisfies $\left(\mathbfcal{T}\right)_{i>j} = 0$, i.e., Eq.~\eqref{eq:transition_matrix} is an upper triangular invertible matrix.
With the definitions in Eq.~\eqref{eq:state_vector} and Eq.~\eqref{eq:transition_matrix} we can rewrite the marginals as
\begin{equation}
\label{eq:matrix_equation}
    \mathbf{P}^{(D)} = \mathbfcal{T}^{D-1} \mathbf{P}^{(1)}.
\end{equation}
Since we start with a root node with $N$ encoded clauses,
\begin{equation}
    P^{(1)}_{i} = \delta_{iN}.
\end{equation}
$\mathbf{P}^{(D)}$ can be obtained by using Eq.~\eqref{eq:matrix_equation} and $C$ can be computed as 
\begin{equation}
    C = K^{D-1} \left( 1 - (\mathbf{P}^{(D)})_0 \right).
\end{equation}
To obtain the explicit expression in Eq.~\eqref{eq:CvsK,D}, we proceed as follows. Define the right eigenvectors of the transition matrix
\begin{align}
\sum_{j = 0}^{N} \mathbfcal{T}_{ij} v_{j}^{\mu} = \lambda^{\mu} v_{i}^{\mu}, \quad \mu = 0, 1, ..., N,
\end{align}
Since $\mathcal{T}$ is upper triangular, the eigenvalues are the just the diagonal elements
\begin{align}
    \lambda^{\mu} = P(\mu | \mu) = \frac{1}{Z_{K}(\mu)}. \label{eq:evals}
\end{align}
The initial state is given by ${\bf P}_{i}^{(1)} = \delta_{i N}$, which we can write as a linear superposition of eigenvectors
\begin{align}
    {\bf P}_{i}^{(1)} = \delta_{i N} = \sum_{\mu = 0}^{N} \omega_{\mu} v_{i}^{\mu}. \label{eq:weight_def}
\end{align}
Using this representation, we can compute the probability distribution for arbitrary depth by using Eq.~\eqref{eq:matrix_equation} to get 
\begin{align}
    {\bf P}_{i}^{(D)} =\sum_{\mu = 0}^{N} \lambda_{\mu}^{D-1} \omega_{\mu}  v_{i}^{\mu}
    \label{eq:p_vec_sum}
\end{align}
From this, we need the zeroth component to compute $C$, giving the representation
\begin{align}
    C = K^{D-1} \left( 1 - \sum_{\mu = 0}^{N}\lambda_{\mu}^{D-1} \omega_{\mu}   v_{0}^{\mu} \right). \label{eq:C_spec}
\end{align}
Now that we have the spectral representation of $C$, what remains is to determine the initial weights $\omega_{\mu}$, and the explicit eigenvectors $v_{i}^{\mu}$. We state the results and prove them below. 
The eigenvectors take the remarkably simple form, {\it independent of $K$}, given in terms of binomial coefficients
\begin{align}
     v_{i}^{0} = \delta_{i 0}, \quad v_{i}^{\mu} = \begin{cases} (-1)^{\mu + i} \left( { \mu \atop i}\right) \,\,0 \le i\le  \mu\\
     0 , \quad i >\mu \end{cases} \quad {\rm for} \quad \mu > 0. \label{eq:evecs}
\end{align}
We display the first few vectors below, since the structure may not be immediately apparent from the formula above:
\begin{align}
    v^{0} &= (1, 0, 0, 0, ..., 0)\\
    v^{1} &= (-1, 1, 0, 0,...., 0)\\
    v^{2} & = (1, -2, 1, 0, 0, ..., 0)\\
    v^{3} & = (-1, 3, -3, 1, 0, ..., 0)\\
    ...
\end{align}
and so on. The initial weights are given by
\begin{align}
\omega_{\mu} = \left( { N \atop \mu}\right). \label{eq:weights}
\end{align}
Using Eqs. \eqref{eq:weights}, \eqref{eq:evecs}, and \eqref{eq:evals} in \eqref{eq:C_spec} recovers Eq.~\eqref{eq:CvsK,D} in the main text. \\

{\bf Proof:} 
First, we prove Eq \eqref{eq:evecs}. It is easy to see by inspection that $v^{0}$ is an eigenvector of the transition matrix with eigenvalue $1$. We may construct the other eigenvectors iteratively by defining a shift operator
\begin{align}
    \left(\mathcal{P}v\right)_{i} =  v_{i-1} .
\end{align}
Then we claim that
\begin{align}
    v^{\mu} \equiv \left( - 1 + \mathcal{P}\right)^{\mu} v^{0}, \label{eq:gen_evec}
\end{align}
is also an eigenvector of the transition matrix, which means
\begin{align}
    \mathbfcal{T} ( - 1 + \mathcal{P})^{\mu} v^{0} = \lambda_{\mu} \left( - 1 + \mathcal{P}\right)^{\mu} v^{0}.
\end{align}
This equation will hold iff 
\begin{align}
    (v^{0})^{T} \left( ( - 1 + \mathcal{P})^{\mu} \right)^{-1} \mathbfcal{T} (- 1 + \mathcal{P})^{\mu} v^{0} = \lambda_{\mu}.
\end{align}
Some algebra shows that the LHS reduces to
\begin{align}
   S_{\mu} \equiv    (v^{0})^{T} \left( ( - 1 + \mathcal{P})^{\mu} \right)^{-1} \mathbfcal{T} (- 1 + \mathcal{P})^{\mu} v^{0} = \sum_{j = 0}^{\mu} (-1 )^{ j} \left( { \mu \atop j}\right) \mathbfcal{T}_{0 j}.
\end{align}
We can use the explicit form of the matrix elements
\begin{align}
    \mathbfcal{T}_{0j} = \frac{K-1}{j + K - 1}.
\end{align}
Next, using the identity
\begin{align}
    \sum_{j = 0}^{\mu} (-1 )^{ j } \left( { \mu \atop j}\right) \frac{1}{j + b} = \frac{(b - 1)! \mu!}{(b + \mu)!},
\end{align}
We get
\begin{align}
    S_{\mu} &= \frac{(K - 1)! \mu!}{(\mu + K - 1)!} = \frac{1}{Z_{K}(\mu)} = P(\mu | \mu) = \lambda_{\mu}
\end{align}
Therefore, we have shown that $S_{\mu} = \lambda^{\mu}$, which confirms that $v^{\mu}$ defined in \eqref{eq:gen_evec} is indeed an eigenvector of the transition matrix with eigenvalue $\lambda^{\mu}$. Next, it is fairly straightforward to see that this relationship implies a recursion relation
\begin{align}
    v^{\mu} = ( - 1 + P) v^{\mu-1}
\end{align}
which in components reads
\begin{align}
    v_{i}^{\mu} = - v_{i}^{\mu-1} + v_{i-1}^{\mu-1},
\end{align}
This is precisely the recursion relation for Pascal's rule, except for the minus sign in front of the first term. That can be compensated for if $v_{i}$ and $v_{i-1}$ have opposite signs. Therefore, the solution must be the binomial coefficient but with alternating signs. Plugging in Eq.~\eqref{eq:evecs} confirms this is the case. 
Finally, we need to show that the weights $\omega_{\mu}$ defined in Eq.~\eqref{eq:weight_def} are given by Eq.~\eqref{eq:weights}. We do this by plugging it in, we have for $i < N$:
\begin{align}
    \sum_{\mu = 0}^{N} \omega_{\mu} v_{i}^{\mu} &= \sum_{\mu = i}^{N} ( - 1)^{i + \mu}\left( { N \atop \mu}\right)  \left( { \mu \atop i}\right)\\
    & = \sum_{\mu = i}^{N} ( - 1)^{i + \mu}\left( { N \atop i}\right)  \left( { N-i \atop \mu - i}\right)\\
    & = \left( { N \atop i}\right)  \sum_{m = 0}^{N - i} ( - 1)^{m} \left( { N - i \atop m}\right) = \left( { N \atop i} \right) ( 1 -1)^{N-i} = 0
\end{align}
However, for $i = N$, we have
\begin{align}
   \sum_{\mu = 0}^{N} \omega_{\mu} v_{N}^{\mu}  = \omega_{N} v_{N}^{N} = 1,
\end{align}
which shows the weighted sum is just the delta function as required per Eq.~\eqref{eq:weight_def}. \\

From Eq.~\eqref{eq:p_vec_sum}, the compression ratio distribution at level-$D$ reads (writing $n^{(D)}:=n$)
\begin{equation}
    P(n) = \sum_{\mu=n}^N \frac{(-1)^{\mu+n}}{Z_K(\mu)^{D-1}} \binom{N}{\mu} \binom{\mu}{n}.
    \label{eq:compression_ratio_formula}
\end{equation}
In particular, Eq.~\eqref{eq:compression_ratio_formula} has an interesting connection with the Riemann zeta function. Choosing $K=2$, we have 
\begin{equation}
    P(n=0) = \sum_{\mu=0}^N \frac{(-1)^{\mu}}{(\mu+1)^{D-1}} \binom{N}{\mu},
    \label{eq:p0_formula}
\end{equation}
which turns out to be the same as the memory retention curve for forgetting model III in \cite{katkov2022mathematical}. 
On the other hand, one of the globally convergent series representation of the Riemann zeta function is \cite{hasse1930summierungsverfahren}
\begin{equation}
    \zeta(D) = \frac{1}{D-1} \sum \limits_{N=0}^{\infty} \frac{1}{N+1} \sum \limits_{\mu=0}^N \frac{(-1)^\mu}{(\mu+1)^{D-1}} \binom{N}{\mu}.
\end{equation}
Analytically continue $D$ to the complex plane in Eq.~\eqref{eq:p0_formula}, we have
\begin{equation}
 \zeta(D) = \frac{1}{D-1} \sum \limits_{N=0}^{\infty} \frac{P(n=0)}{N+1} .
\end{equation}
Now using the definition of recall length $C(N)=K^{D-1}[1-P(n=0)]$ for $K=2$, and identify $\gamma(N) = C(N)/(N+1)$ as the global compression ratio for the entire recall, we have the following identity
\begin{equation}
    \sum_{N=0}^{\infty} \gamma (N) = 2^{D-1}\big[H_\infty - (D-1)\zeta(D)\big], 
    \label{eq:zeta_relation}
\end{equation}
where $H_\infty$ is the Harmonic series $H_\infty = \sum_{N=1}^{\infty}1/N$. Eq.~\eqref{eq:zeta_relation} then provides a connection between the average global compression ratio of narrative recall with the Riemann zeta function.

\subsubsection*{Asymptotic scale-invariant compression ratios distribution}

In the limit of large $n$, and assuming $n\gg K$, we can use Stirling's approximation to write
\begin{equation}
\log Z_K(n) = \log \frac{n^{K-1}}{(K-1)!} + O\left(\log n\right).
\end{equation}
Then we can rewrite the conditional probability given by Eq.~\eqref{eq:conditional} in terms of the normalized compression ratios $s_{l} = n^{(l)}/N$:
\begin{align}
    \rho(s_{l+1}| s_{l}) \equiv N P(N s^{(l+1)}| N s^{(l)}) =N \frac{Z_{K-1}(N (s_{l} - s_{l+1}))}{Z_{K}(N s_{l})} \approx (K-1)\frac{1}{s_{l}} \left( 1 - \frac{s_{l+1}}{s_{l}}\right)^{K-2}.
\end{align}

\subsection{Experimental data collection}

In our experiments, we used a set of 11 narratives. The narratives are based on the oral retelling of personal experiences told by real people, as presented in \cite{labov2013language,johnstone1990stories}, and variants of those generated by GPT-4. Original narratives were corrected for spelling and clarity. 8 of the narratives and the corresponding recall data were taken from our previous publication \cite{georgiou2023using}. We performed additional recall experiments on 3 new narratives of similar style with $L=81,122,194$, two taken from \cite{johnstone1990stories} and the longest one from \cite{labov2013language} (reproduced in the section below). We followed the experimental protocol in \cite{georgiou2023using}: for each narrative, 100 subjects were recruited from the online crowd-sourcing platform called Prolific (www.prolific.com), and presented with the narratives on a computer screen with rolling text. After the presentation of the narrative, the subjects were asked to perform a free recall experiment and write down their recall of the narrative, including as many details as possible. 

After collecting the recalls, each recall is segmented into clauses by GPT-4 using the following prompt:

\begin{prompt}[prompt:seg_fine]{Recall Segmentation}
\texttt{Provide a word-for-word segmentation of the following narrative into linguistic clauses, numbered in order of appearance in the narrative: \{recall of a narrative\}}
\end{prompt}

where the \texttt{\{recall of a narrative\}} box is a placeholder for the recall from the subjects. An example output of a segmented recall is provided in Section~\ref{appsec:recall}.

\subsection{Recall mapping and compression ratios}
\label{appsec:recall_mapping}

\begin{figure*}[ht]
    \centering
    \includegraphics[width=0.6\linewidth]{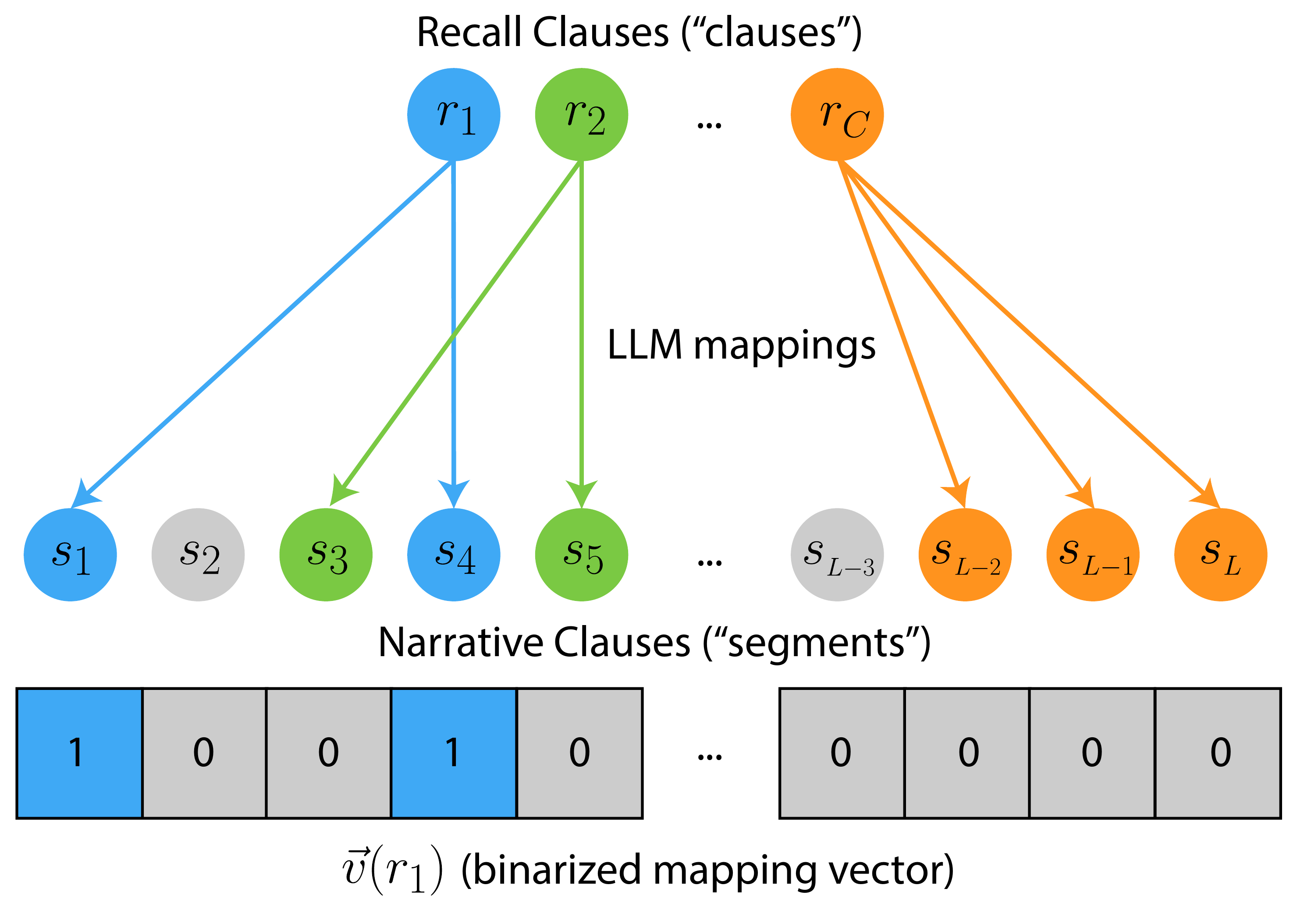}
    \caption{{\bf Schematics for recall mappings.} Top: $C$ recall clauses (referred to as ``clauses'' in the prompts). Middle: $L$ narrative clauses (referred to as ``segments'' in the prompts). Arrows indicate mappings generated by LLMs. Bottom: An example binarized mapping vector $\Vec{v}$ for the first recalled clause $r_1$, where mapped narrative clauses are assigned a value of 1, and unmapped ones are assigned a value of 0. }
    \vspace{-1em}
    \label{appfig:mapping}
\end{figure*}

We utilized zero-shot prompting with OpenAI's model \texttt{gpt-4-turbo-2024-04-09} to project each individual recall back into the corresponding narrative, as shown schematically in Fig.~\ref{appfig:mapping}. For a given recall clause, the compression ratio $n$ is defined as the number of narrative clauses it is projected into. Recall clauses that do not get any projections are considered to be mistaken intrusions, with their overall fraction being $2.7\%$. They are not included in the calculation of recall length presented in Fig.~\ref{fig:data_figure}(b). Self-referencing mistakes, where a recall clause is mistakenly mapped to the segment with the same serial position (segment number), can also occur in the LLM-generated mappings. As in Section~\ref{appsec:completion} for clause number 20 that is mistakenly mapped to segment number 20. However, our analysis found that this happens rarely, accounting for only $1\%$ of all mapped clauses for the longest narrative ($L=194$) and $0.3\%$ for the second longest ($L=130$).

The recall mapping prompt, Prompt~\ref{prompt:rec_map}, is a text block that takes the recall clauses and narrative segments as arguments. After specifying the recall clauses and narrative segments (e.g., Sections~\ref{appsec:narr_segs}-\ref{appsec:recall}), Prompt~\ref{prompt:rec_map} is given to GPT-4, which outputs the mapping (e.g., Section~\ref{appsec:completion}). 

\begin{prompt}[prompt:rec_map]{Recall Mapping}
\subsubsection*{System Prompt}

\texttt{You are an AI designed to map a human recall split into clauses and an original narrative split into segments, both provided in JSON format. Your task is to generate a JSON output that maps the clauses to the relevant segments. Clauses can be mapped to multiple segments, and the same segment can be mapped multiple times. If a recall clause cannot be mapped to any narrative segment, return an empty list. Here is an example of the desired output format:}

\begin{lstlisting}[language=json,firstnumber=1]
{"mappings": 
   [{"clause": 1, "segments": [1]},
    {"clause": 2, "segments": [3, 7]},
    {"clause": 3, "segments": []},
    {"clause": 4, "segments": [2, 4, 5, 7]}]
}
\end{lstlisting}

\subsubsection*{User Prompt}
Input JSON for Recall Clauses:

\begin{lstlisting}[language=json]
{clauses}
\end{lstlisting}

Input JSON for Original Narrative Segments:

\begin{lstlisting}[language=json]
{segments}
\end{lstlisting}

\end{prompt}

\subsection{Example narrative ($L=194$)}

\subsubsection{Segmented clauses from the narrative}
\label{appsec:narr_segs}
\subsubsection*{Originally from Mary Costa: “The death of her youngest daughter” \cite{labov2013language}}

\begin{lstlisting}[language=json,firstnumber=1]
{"segments": {
    "1": "My daughter died.",
    "2": "It'll be twenty-eight years this month.",
    "3": "The thirtieth of this month, yeah.",
    "4": "She died at Methodist Hospital.",
    "5": "She had an operation for appendicitis.",
    "6": "It was late Saturday afternoon when they operated, very late.",
    "7": "Sunday morning, my oldest daughter, Victoria, went to the hospital to see how she was.",
    "8": "She was still under ether.",
    "9": "Back then, they used a lot of ether.",
    "10": "There was a chair there.",
    "11": "They were just starting to get people out of bed, you know.",
    "12": "So, Victoria saw the chair,",
    "13": "and she said to the nurse, 'What's this chair doing here?'",
    "14": "The nurse said, 'Well, we've got to get her out of bed.'",
    "15": "She said, 'What are you talking about? She's still under ether.'",
    "16": "'She's too sick.'",
    "17": "But the nurse said, 'Well, that's the doctor's orders.'",
    "18": "So, they went to get her out of bed,",
    "19": "and she passed out,",
    "20": "and they had to put her back in bed."
    "21": "See, here, she never - I don't know if they knew or what,",
    "22": "but I blame them for the blood clot she got, taking her out of bed.",
    "23": "So, when she was walking around to come home after ten days,",
    "24": "I was waiting for the call from the hospital.",
    "25": "I didn't have a phone then.",
    "26": "The lady from the store who had a phone called me, 'Mrs. Costa,' at a quarter to five at night.",
    "27": "'Mrs. Costa, they just called you from the hospital.",
    "28": "Go pick Marie up.",
    "29": "She's discharged.'",
    "30": "I had all the clothes laid out on the table.",
    "31": "And Rita was - I was cooking,",
    "32": "I said, 'Rita, finish the cooking.'",
    "33": "'Daddy will be home soon.'",
    "34": "He was working down in the yard.",
    "35": "I said, 'And put a big pot of coffee on,' because she had been wishing for coffee.",
    "36": "And you know, every morning, that first cup of coffee, I offered it to her.",
    "37": "When I get up in the morning, the first cup is hers.",
    "38": "Since she's gone, I never forget that.",
    "39": "So I said, 'Put a pot of coffee on,",
    "40": "and send for some cake from the bakery.'"
    "41": "I'm going to get Marie; they called me."",
    "42": ""All right, mother, go ahead," she said, "I'll finish up."",
    "43": "And I went myself. You know, I had the coat on my arm.",
    "44": "She was operated on in March -",
    "45": "No, April! So that was the thirtieth.",
    "46": "She was operated - ten days before, so that would be the twentieth.",
    "47": "The twentieth of April, right? It was kind of chilly.",
    "48": "Oh, and I had everything in a bag,",
    "49": "and I had the coat on my arm and her dress hanging on my arm.",
    "50": "She saw me from the windows, from Wolf Street.",
    "51": "She said, "My mother's just coming."",
    "52": "She was with the girls, saying goodbye to everyone - you know how it is.",
    "53": "They told me on the phone, it would be ten dollars more for her board.",
    "54": "And to make sure I brought her girdle to put on.",
    "55": "I said, "All right, I'll be right over."",
    "56": "And I went.",
    "57": "But when I got there, the nurse said to me, "Sit down, Mrs. Costa, sit here for a while."",
    "58": "So I sat down, and I was praying for another lady in the hospital that had an operation for her gall bladder.",
    "59": "She lived near my sister, and she had ten children.",
    "60": "They had said she was in very bad shape."
    "61": "I prayed, "Oh dear God, don't take that mother away from her children."",
    "62": "I was praying for her, you know.",
    "63": "Then a little nurse's aide went to the closet and got some blankets.",
    "64": "I started getting the chills,",
    "65": "and I was thinking about it,",
    "66": "when the aide said to the nurse, "Oh, she's going into a cold sweat now."",
    "67": "The nurse hung up the phone and came to me. She said, "Mrs. Costa, has your daughter ever had a fainting spell?"",
    "68": ""Oh my God!" I said, "Don't tell me you're running for my daughter!"",
    "69": ""Yes," she said, "come on in."",
    "70": "So, I went into the room.",
    "71": "And my daughter was on the bed with her bedroom slippers and her housecoat on.",
    "72": "And the beads of sweat - honest to God, not to exaggerate, they were that big.",
    "73": "It came - just poured from her -",
    "74": "I said, "What did you do, kill my daughter?",
    "75": "You killed her! You killed her!"",
    "76": "That's what I kept yelling.",
    "77": "And the doctor rushed in,",
    "78": "see, they were trying to get the doctor, you know.",
    "79": "She was a lady doctor.",
    "80": "Dr. Schwartz actually did the operation, but she was his assistant, Dr. Montique."
    "81": "She was a lady doctor.",
    "82": "So, she came in and she said -",
    "83": "I said, "You killed my daughter - what did you do to her?"",
    "84": "She said, "We don't know what happened, Mrs. Costa."",
    "85": "She took me away from the bed.",
    "86": "The nurse said to me, "Talk to her."",
    "87": "I said, "Talk to her? My daughter's dying, what am I supposed to tell her?"",
    "88": "And I was screaming.",
    "89": "My hair was all pulled out.",
    "90": "I went hysterical, you know what I mean? She was only eighteen years old.",
    "91": "Beautiful girl, you can see from her picture.",
    "92": "And just as I turned my back, my daughter passed away.",
    "93": "She was getting ready to go home.",
    "94": "I had the clothes and everything ready.",
    "95": "My husband was waiting for her.",
    "96": "He came home, went upstairs to shave and wash, then sat on the step, waiting.",
    "97": ""When is the mother coming?"",
    "98": "The hospital kept calling.",
    "99": "The people in the office didn't know what had happened yet.",
    "100": "They kept calling the store, trying to reach us."
    "101": "They asked, "When is the mother coming to pick up Marie?",
    "102": "She's discharged."",
    "103": "And Rita said "That's strange, the mother went to the hospital. She should be there."",
    "104": "So they came back to the hospital to look for us.",
    "105": "Rita said, "Gee, that's strange,"",
    "106": "she said, "Mother must've gotten really worked up about Marie."",
    "107": "You know, I've always had a bad heart.",
    "108": ""Mother must've gotten so upset about Marie, she probably made herself sick," Rita thought.",
    "109": "So she came over with my son-in-law, in the car.",
    "110": "And I was in the main hallway, at the office.",
    "111": "I was all pulled apart.",
    "112": "One doctor wanted to give me a shot.",
    "113": "I said, "Get out of here,",
    "114": "I'll kick you," I said.",
    "115": ""You killed my daughter, now you want to kill me?"",
    "116": "Another nurse came over.",
    "117": "See, they all knew me.",
    "118": "I had three operations in no time, not even a year.",
    "119": "And I knew all the nurses.",
    "120": ""Mrs. Costa, please, take -""
    "121": ""Get out!" I said, "Don't you dare come near me."",
    "122": "I was so hysterical.",
    "123": "I was banging my head against the wall and everything.",
    "124": "They had to grab me.",
    "125": "Then they brought me to the office, see.",
    "126": "When Rita came in and saw me like that, she asked, "Mother, what happened?"",
    "127": "I said, "We lost Marie.",
    "128": "Marie died."",
    "129": "Well, that's this one. This is a picture of Rita.",
    "130": "She went down like lightning.",
    "131": "Dropped to the floor, and that's a marble floor.",
    "132": "She went down like lightning.",
    "133": "And the back of her leg, it felt just like a pole.",
    "134": "Hard like a pole, the back of her leg.",
    "135": "And that girl is suffering.",
    "136": "Thirty - no, twenty-eight years since her sister died?",
    "137": "When she gets that pain, it's like she has phlebitis from it.",
    "138": "She suffers - seven, eight, nine weeks at a time.",
    "139": "She can't walk, she goes through such pain since then.",
    "140": "And how did they figure how Marie died?"
    "141": "The girl who was walking with Marie,",
    "142": "told the doctor that Marie was saying goodbye to Arlene.",
    "143": "And all Marie said to that girl was, "Oh, I've got a pain in my leg and I can't see."",
    "144": "And then Marie collapsed.",
    "145": "Later, they said it was a blood clot.",
    "146": "I said, "Don't you dare touch my daughter.",
    "147": "I'll burn this hospital down!",
    "148": "I don't care who's in it."",
    "149": "I said, "Don't you dare touch my daughter.",
    "150": "You want to find out what she died from.",
    "151": "But if you touch my daughter, I'll kill you."",
    "152": "I wouldn't let them touch her.",
    "153": "So when I got home - they took me home.",
    "154": "I was screaming in the car.",
    "155": "I couldn't get out, you know, I couldn't -",
    "156": "I was screaming!",
    "157": "Everybody came out, coming out of their doors because they could hear the screams.",
    "158": "My husband was sitting on the step.",
    "159": "He asked, "What's the matter? What happened to you?"",
    "160": "Well, my hair was all pulled, and my face - I was a mess."
    "161": "I didn't know what I had done to myself.",
    "162": "I said, "Mike, we lost Marie. Marie died." It was just a shock.",
    "163": "My husband had a heart attack.",
    "164": "Yes, he had a heart attack.",
    "165": "Then my son, Michael, yeah. That night.",
    "166": "Right there, my Michael came, and my Michael collapsed.",
    "167": "He had a heart attack too, no kidding.",
    "168": "It was all - my two sisters were here, waiting until about half-past four.",
    "169": "They waited because they knew she was coming home that day, waiting for the call.",
    "170": "My younger sister had brought an orange for Marie, and I had never seen an orange like it. It was a big orange, like that.",
    "171": "She said, "Mary, look, when they call you, when you pick Marie up, let us know, and we'll come over."",
    "172": "I said, "All right."",
    "173": "So, I got home and called everybody, told them what happened.",
    "174": "I called my brother-in-law.",
    "175": "He was an undertaker.",
    "176": "My sister's husband - he's passed away now too. God rest his soul.",
    "177": "Maybe you know him...",
    "178": "So they were waiting for my call, you know,",
    "179": "but instead, I sent my son James, he was little then.",
    "180": "I said, "Go tell Aunt Millie and Aunt Jenny.""
    "181": "They had cleaned the pavement,",
    "182": "and done the cement work, and they were cleaning the pavement.",
    "183": "They said, "Here comes little Jamie," and now they said,",
    "184": ""Maybe he's come to tell us Marie's home."",
    "185": "But when he got there, the poor kid's eyes were all red.",
    "186": "They said, "James, what's the matter, honey?",
    "187": "Why are you crying?"",
    "188": ""Aunt Jenny, Aunt Millie, Marie died," he said.",
    "189": "Well, my sisters and my mother, you know, they were all alive.",
    "190": "It was a shock to everybody. That was a shock that -",
    "191": "The neighborhood - Annie at the store always says, "Mrs. Costa," every time I go in there, she reminds me, "I just picture Marie standing at that step, looking over here." She used to stand at that step.",
    "192": "She was a real quiet girl, pleasant, didn't talk much,",
    "193": "but she was so good, I can't even tell you.",
    "194": "That's all. So that's what happened."
    }
}
\end{lstlisting}

\subsubsection{Segmented clauses from an individual recall}
\label{appsec:recall}
\begin{lstlisting}[language=json,firstnumber=1]
{"clauses": {
    "1": "It had been 28 years since the narrator's daughter, Marie, passed away.",
    "2": "Marie had been operated on for appendicitis at Methodist Hospital.",
    "3": "After the surgery, they used a lot of ether.",
    "4": "When the nurse tried to get Marie out of bed, she passed out.",
    "5": "She had to be put back in bed.",
    "6": "The narrator blamed the hospital staff for a blood clot Marie developed from being moved too soon.",
    "7": "After 10 days, the hospital called to say Marie was discharged.",
    "8": "The narrator went to pick her up.",
    "9": "She brought clothes and coffee for Marie.",
    "10": "Marie had been wishing for it.",
    "11": "However, when the narrator arrived, a nurse told her to sit down and wait.",
    "12": "The narrator prayed for another patient.",
    "13": "Then a nurse informed her that Marie was sweating heavily and appeared to be in distress.",
    "14": "The narrator rushed into the room.",
    "15": "She saw Marie sweating profusely.",
    "16": "In a panic, she accused the doctors and nurses of killing her daughter.",
    "17": "Soon after, Marie passed away while preparing to come home.",
    "18": "The hospital staff had tried to contact the family.",
    "19": "When the narrator's other daughter, Rita, and son-in-law arrived, they found the narrator in a state of hysterics.",
    "20": "The news was devastating.",
    "21": "It led to more health complications for family members."
}
\end{lstlisting}

\subsubsection{LLM output}
\label{appsec:completion}
\begin{lstlisting}[language=json,firstnumber=1]
{"mappings": 
   [{"clause": 1, "segments": [1, 2},
    {"clause": 2, "segments": [4, 5]},
    {"clause": 3, "segments": [8, 9]},
    {"clause": 4, "segments": [11, 14, 18, 19]},
    {"clause": 5, "segments": [20]},
    {"clause": 6, "segments": [21, 22]},
    {"clause": 7, "segments": [23, 24, 26, 27, 28]},
    {"clause": 8, "segments": [43, 56]},
    {"clause": 9,  "segments": [30, 31, 35, 39, 40, 41]},
    {"clause": 10, "segments": [35]}
    {"clause": 11, "segments": [57]},
    {"clause": 12, "segments": [58, 59, 60, 61]},
    {"clause": 13, "segments": [66, 67]},
    {"clause": 14, "segments": [70]},
    {"clause": 15, "segments": [71, 72, 73]},
    {"clause": 16, "segments": [74, 75, 76, 83, 146, 147, 148, 149, 150, 151]},
    {"clause": 17, "segments": [93]},
    {"clause": 18, "segments": [98, 99, 100]},
    {"clause": 19,  "segments": [104, 105, 106, 109, 110, 111, 112, 113, 114, 115, 116, 117, 118, 119, 120, 121, 122, 123, 124, 125, 126]},
    {"clause": 20, "segments": [20]}
    {"clause": 21, "segments": [107, 108, 133, 134, 135, 136, 137, 138, 139, 163, 164, 165, 166]}
}

\end{lstlisting}

\subsection{Validation with other LLM-generated Mappings}
\begin{figure*}[ht]
    \centering
    \includegraphics[width=0.9\linewidth]{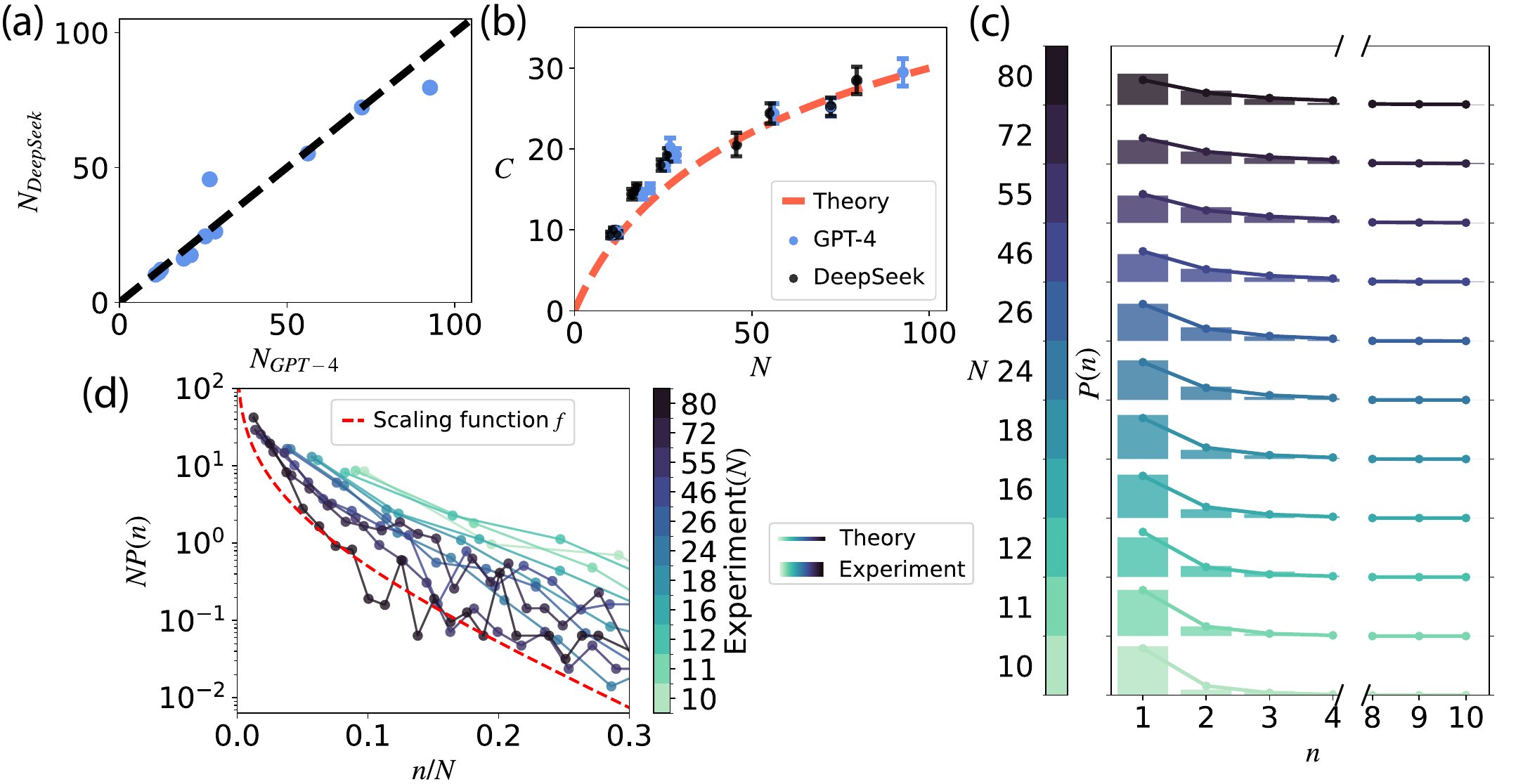}
    \caption{{\bf Comparison between two language models} \textbf{(a)} Average size of the tree memory representation of each narrative ($N$) generated by GPT-4 vs DeepSeek-V3. The dashed line corresponds to the diagonal. \textbf{(b)} The mean number of recalled clauses $C$ vs. average $N$, for all 11 narratives. Blue filled circles - GPT-4 generated mappings. Black filled circles - DeepSeek-V3 generated mappings. Red dashed line - theoretical prediction for $K=D=4$, same as in the main text, Fig.~2(b). Error bars in (a,b) are standard error of the mean. \textbf{(c)} Normalized empirical histograms of compression ratios for all subjects separately for each narrative, as measured from DeepSeek-V3-generated mappings. Solid lines - theoretical predictions obtained from $K=D=4$, same as in the main text, Fig.~2(c). The range between tick marks on the y-axis is $[0,1]$. {\bf (d)} The distribution of experimentally measured compression ratios relative to $N$ as mapped by DeepSeek-V3. The scaling function $f$ is the same as in the main text, Fig.~2(d). }
    \vspace{-1em}
    \label{appfig:deepseek}
\end{figure*}

To evaluate the robustness of our results, we repeat our analysis using the above mappings obtained with another LLM, the recently released open-source model DeepSeek-V3 \cite{liu2024deepseek}, following the same procedure as in Sec.~\ref{appsec:recall_mapping} and the same Prompt~\ref{prompt:rec_map}. As shown in Fig.~\ref{appfig:deepseek}, the summary statistics of the recall obtained with DeepSeek are quite similar to those obtained with GPT-4 and hence align with model predictions. Note that using DeepSeek results in slightly different values of $C$ for some of the narratives because some of the recall clauses are not mapped to any of the narrative clauses by one model but are mapped by the other. 

\begin{figure*}[ht]
    \centering
    \includegraphics[width=0.9\linewidth]{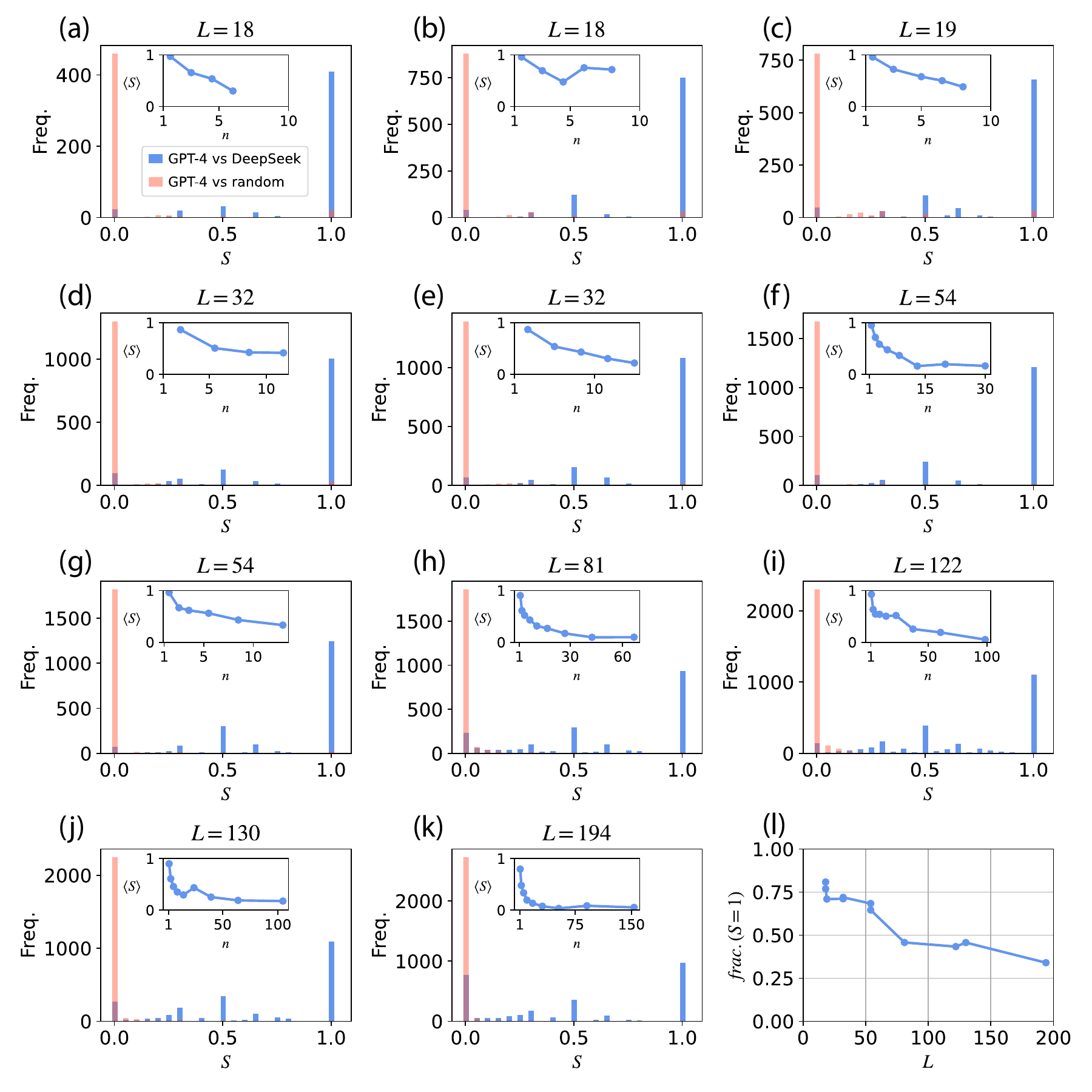}
    \caption{{\bf Similarity between mappings generated by GPT-4 and DeepSeek-V3.} \textbf{(a-k)} Distribution of normalized similarity scores $S$ for each mapped recall clauses across the 11 narratives analyzed in the main text. The inset shows the averaged similarity score $\langle S \rangle$ within each bin versus compression ratios $n$, where the bins are chosen uniformly on a linear scale for (a)-(e) and uniformly on a logarithmically scale for (f)-(k). \textbf{(l)} The fraction of recall clauses with a perfect maximum similarity score between the two mappings ($S=1$) vs. narrative length $L$. }
    \vspace{-1em}
    \label{appfig:overlap}
\end{figure*}

To evaluate the similarity between the mappings produced by the two LLMs at the level of individual clauses, we used binary mapping vectors for each recall clause (see bottom of Fig.~\ref{appfig:mapping}). Denoting these vectors as $\Vec{v}(r_i)$ and $\Vec{w}(r_i)$, respectively for GPT and DeepSeek, we computed the corresponding similarity score as the normalized dot product of these two vectors (also known as the Jaccard Similarity):
\begin{equation}
S\bigl(\vec{v}(r_i),\vec{w}(r_i)\bigr) = \frac{\vec{v}(r_i) \cdot \vec{w}(r_i)}{\|\vec{v}(r_i)\|_1 + \|\vec{w}(r_i)\|_1 - \vec{v}(r_i) \cdot \vec{w}(r_i)},
\label{appeq:similarity}
\end{equation}
where $\|\cdot \|_1$ is the $1$-norm. The denominator in Eq.~\eqref{appeq:similarity} is the union of the non-zero entries from the two mapping vectors. Defined in this way, the similarity score ($S$) equals 1 when the two vectors are identical (perfect overlap) and 0 when there is no overlap. In Fig.~\ref{appfig:overlap} (a)-(k), we show the distribution of $S$ over all recall clauses from $100$ subjects for each of the 11 narratives used in this study (shown in blue). For comparison, we also shuffled the mapping vectors produced by DeepSeek-V3, preserving the number of nonzero elements in each vector  (referred to as ``random mapping") and calculated its similarity score with the GPT-4 mapping vectors (shown in red). Most of the mapping vectors agree perfectly ($S=1$) between GPT-4 and DeepSeek-V3, whereas the similarity score between the mapping vectors of GPT-4 and shuffled DeepSeek-V3 is very low. Fig.~\ref{appfig:overlap} (l) further shows the fraction of recall clauses with perfect similarity between the models (corresponding to the last bin in (a)-(k)) as a function of narrative length $L$. This fraction decreases with $L$, as recall becomes more summarizing and abstract in nature. To better understand what causes the decrease in the overlaps, in the inset of Fig.~\ref{appfig:overlap} (a)-(k) we bin the scores based on the maximum of the compression ratios $n$ of the two mapping vectors, and show the averaged similarity score $\langle S\rangle$ within each bin as a function of $n$. In general, we observe a decrease of $\langle S\rangle$ with larger $n$, indicating that for highly summarizing recall clauses, there is little agreement between the two models. We believe that these differences in mappings reflect variations in architectural details, training datasets and fine-tuning algorithms of the models, which may therefore lead them to ``comprehend" the narrative and its recalls differently, just as different humans would. 

\end{document}